\documentclass[aps,prx,twocolumn,floatfix,longbibliography,nofootinbib,superscriptaddress]{revtex4-2}
\usepackage[utf8]{inputenc}
\usepackage{natbib}
\usepackage{graphicx}
\usepackage{xcolor}
\usepackage[tbtags]{amsmath}
\usepackage[colorlinks, linkcolor=blue]{hyperref}
\hypersetup{colorlinks,allcolors=black}
\usepackage{amssymb}
\usepackage{amsmath}
\usepackage{float}
\usepackage{amsmath}
\usepackage{tabularx,graphicx}
\usepackage{epstopdf}
\usepackage{latexsym}
\usepackage{color, colortbl}
\usepackage{psfrag}
\usepackage{bbm}
\usepackage{bm}
\usepackage{titlesec}
\usepackage{dsfont}
\usepackage{feynmp}
\usepackage{slashed}
\usepackage{multirow}
\usepackage[normalem]{ulem}
\usepackage{subfigure}

\newcommand{\bcen}{\begin{center}}
\newcommand{\ecen}{\end{center}}
\newcommand{\btab}{\begin{tabular}}
\newcommand{\etab}{\end{tabular}}
\newcommand{\bdes}{\begin{description}}
\newcommand{\edes}{\end{description}}

\newcommand{\beq}{\begin{equation}}
\newcommand{\eeq}{\end{equation}}
\newcommand{\bea}{\begin{eqnarray}}
\newcommand{\eea}{\end{eqnarray}}

\newcommand{\bary}{\begin{array}}
\newcommand{\eary}{\end{array}}
\newcommand{\benum}{\begin{enumerate}}
\newcommand{\eenum}{\end{enumerate}}
\newcommand{\bitem}{\begin{itemize}}
\newcommand{\eitem}{\end{itemize}}

\newcommand{\Fig}[1]{Fig.~\ref{#1}}

\makeatletter

\newcommand{\Rmnum}[1]{\expandafter\@slowromancap\romannumeral #1@}
\makeatother

\newcommand{\Li}{\rm{Li}}
\newcommand{\Tr}{\rm{Tr}}
\newcommand{\im}{\rm{Im}}
\newcommand{\real}{\rm{Re}}

\begin{document}

\title{Percolation Transition in a Topological Phase}

\author{Saikat Mondal}
\email{msaikat@iitk.ac.in}
\affiliation{Department of Physics, Indian Institute of Technology Kanpur, Kalyanpur, UP 208016, India}

\author{Subrata Pachhal}
\email{pachhal@iitk.ac.in}
\affiliation{Department of Physics, Indian Institute of Technology Kanpur, Kalyanpur, UP 208016, India}

\author{Adhip Agarwala}
\email{adhip@iitk.ac.in}
\affiliation{Department of Physics, Indian Institute of Technology Kanpur, Kalyanpur, UP 208016, India}

\begin{abstract}
Transition out of a topological phase is typically characterized by discontinuous changes in topological invariants along with bulk gap closings. However, as a clean system is geometrically punctured, it is natural to ask the fate of an underlying topological phase. To understand this physics  we introduce and study both short and long-ranged toy models where a one dimensional topological phase is subjected to bond percolation protocols. We find that non-trivial boundary phenomena follow competing energy scales even while global topological response is governed via geometrical properties of the percolated lattice.  Using numerical, analytical and appropriate mean-field studies we uncover the rich phenomenology and the various cross-over regimes of these systems. In particular, we discuss emergence of ``fractured topological region" where an overall trivial system contains macroscopic number of topological clusters. Our study shows the interesting physics that can arise from an interplay of geometrical disorder within a topological phase. 
\end{abstract}

\maketitle

{\it{Introduction:}}
Percolation transition~\cite{hammersley_1957,priest_1976,reynolds_1977,essam_1980,stauffer_1992,christensen_2002,zallen_2008} is a geometrical phase transition which deals with the connectivity of a graph without any underlying energetic considerations. For short-ranged bond percolation in one dimension~\cite{reynolds_1977,  christensen_2002}, with $p$ as the probability of having the nearest-neighbor bond, the percolation transition happens at $p_{c}=1$. However, in higher dimensions~\cite{Sykes_JMP_1964, Kesten_Springer_1980, Galam_PRE_1996} it is known that $p_c$ can be less than $1$. Critical exponents at the percolation transition follow universal behavior similar to second-order phase transitions~\cite{Kesten_CMP_1987, stauffer_1992,christensen_2002}. Thus,  percolation transitions offer an alternate playground to study critical phenomena which are not engineered by thermal fluctuations but rather are governed by geometrical ``fluctuations" of the graph. Interplay of such geometrical phase transitions with conventional Hamiltonian based classical and quantum symmetry broken phases have been long investigated ~\cite{balian1979ill,aizenman1988,fisher_prl1992,senthil_1996,fisher_physica99,dutta_prb2002,dutta_physicaa2003,dutta_prb2007,divakaran2007,Kovacs_PRR_22}. In particular, the effective dimension of the critical cluster at the percolation threshold has been shown to determine the nature of the phase diagrams both in higher dimensions~\cite{senthil_1996} and in long-range systems even in one dimension \cite{zhang_1983,schulman_1983,schulman_1986,glumac_1993,dutta_physicaa2003,gori17}.  However, similar questions in the context of topological quantum phases have been little explored~\cite{Sahlberg_PRR_2020, Ivaki_PRR_2020}. Such questions can help shed light on a plethora of experimental studies where the interplay of disorder and topological phases have become increasingly important~\cite{
meier2018observation,stutzer2018photonic,
zangeneh2020disorder, Longhi_Optica_20,
kanungo2022realizing,de2019observation,LiPRL2021,meier_2016,shi_2021,thatcher_2022, Gao_PRR_22, Cheng_nature_23}. 

\begin{figure}
    \centering
\includegraphics[width=1.0\columnwidth]{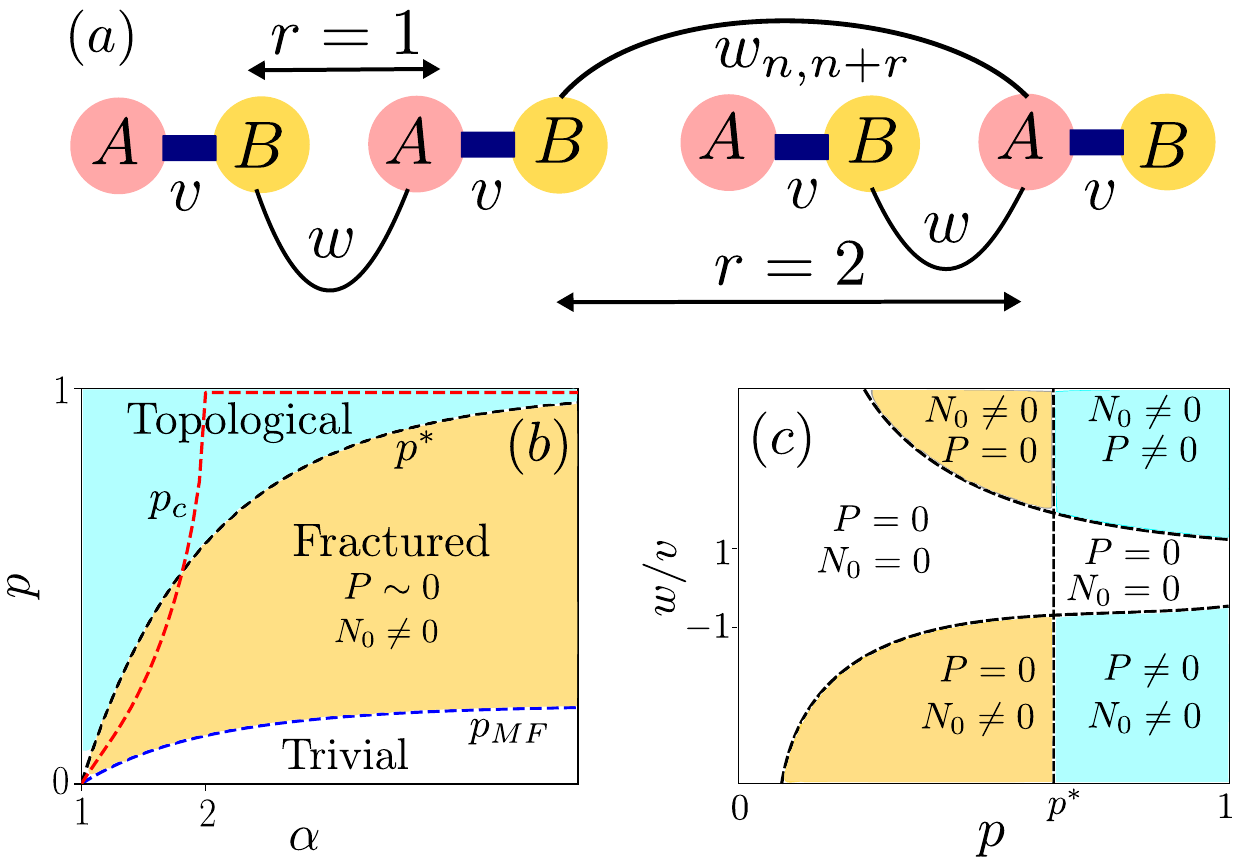}
    \caption{(a) Bond-diluted long-range (LR) SSH chain where intra unit-cell ($A$ and $B$ sites) hopping is $v$.  The probability of having the bond between the unit cells separated by a distance $r$ is $p_{n,n+r}=p/r^{\alpha}$ and the hopping strength is $w_{n,n+r}=w$. In (b), a schematic phase diagram for $p$ and $\alpha$ in the deep topological phase ($|w| \gg |v|$) is shown. Schematic phase diagram for $w/v$ with $p$ for a fixed $\alpha$ is shown in (c), where $p^{*}$ is the threshold where end-to-end connectivity of the chain is established. 
    $P$ and $N_{0}$ are the polarization and the number of zero-energy modes, respectively. The dotted curves in (c) are the mean-field critical curves (see text).}
    \label{fig:1}
\end{figure}

In this work, we pose the question-- can a topological phase and its response functions detect an underlying geometrical (percolation) phase transition? To answer this question we introduce and study two toy models in one dimension where the graph is a one-dimensional chain with short and long range couplings and the decorated Hamiltonian on the graph is the paradigmatic Su-Schrieffer-Heeger (SSH) chain~\cite{ssh_1979, ssh_1980} (see \Fig{fig:1}(a)). The first model we discuss is the short-ranged (SR) model, where the inter unit-cell (composed of sites A and B) bond hopping ($w$) in the conventional SSH model is percolated while the intra unit-cell hopping ($v$) is kept intact. Here, we find interesting signatures of the percolation transition as $p\rightarrow (p_c =1)$ where response function such as polarization~\cite{zak_1989,resta_1998, KANE_2013} approaches $1$ and the system hosts zero energy modes.  
For $p<1$ we show the breakdown of bulk-boundary correspondence where polarization is $\sim${\it zero} even though the system hosts robust zero modes. In order to capture the physics of $p_c<1$ we introduce a long-range (LR) model where inter unit-cell bonds are now long-ranged and chosen probabilistically.  In particular, the probability of a LR bond falls as $p/r^\alpha$ ($\alpha>1$) but the hopping strength is kept as $w$ such that geometrical $p_c<1$. Interestingly, we find that typical signatures of a topological phase in one dimension such as polarization ($P$) and robust zero-modes ($N_{0}$) follow distinctive behaviors reflecting either geometrical or energetic considerations. Using a combination of numerical and ``mean-field" analytic studies, we show that these models can indeed capture interesting aspects of percolation transitions in a topological phase.

{\it{Bond-diluted SSH chain:}} 
We first consider a percolating SR-SSH chain~\cite{ssh_1979,ssh_1980} with the Hamiltonian
\begin{multline}\label{eq_Hsr}
H=v \sum_{n=1}^{L} (c^{\dagger}_{n,A} c_{n,B} + c^{\dagger}_{n,B} c_{n,A}) \\ + \sum_{n=1}^{L-1} w_{n,n+1} (c^{\dagger}_{n,B} c_{n+1,A} + c^{\dagger}_{n+1,A} c_{n,B}), 
\end{multline}
where $L$ is the number of unit cells, $c_{n,A}$ ($c^{\dagger}_{n,B}$) is the annihilation (creation) operator for a fermion in sub-lattice $A$ ($B$) of $n$-th unit cell. The parameters $v$  and $w_{n,n+1}$ denote the intra and inter unit-cell hopping strengths where the latter is chosen from a probability distribution: 
\begin{equation}\label{eq_psr}
{\tilde{\mathcal{P}}}_{\rm{SR}} (w_{n,n+1})=p \delta (w_{n,n+1}-w) + \left(1- p \right) \delta (w_{n,n+1}). 
\end{equation}
Thus, $p$ (where $0 \leq p \leq 1$) is the probability of having a bond between the nearest neighboring unit cells with hopping strength $w$. Choosing $w$ and $v$ as real, our system always belongs to the symmetry class BDI~\cite{altland_1997,adhip_shenoy_2017}. We always study the system at half-filling, such that Fermi energy remains pinned to zero.

\begin{figure}
    \centering
\includegraphics[width=1.0\columnwidth]{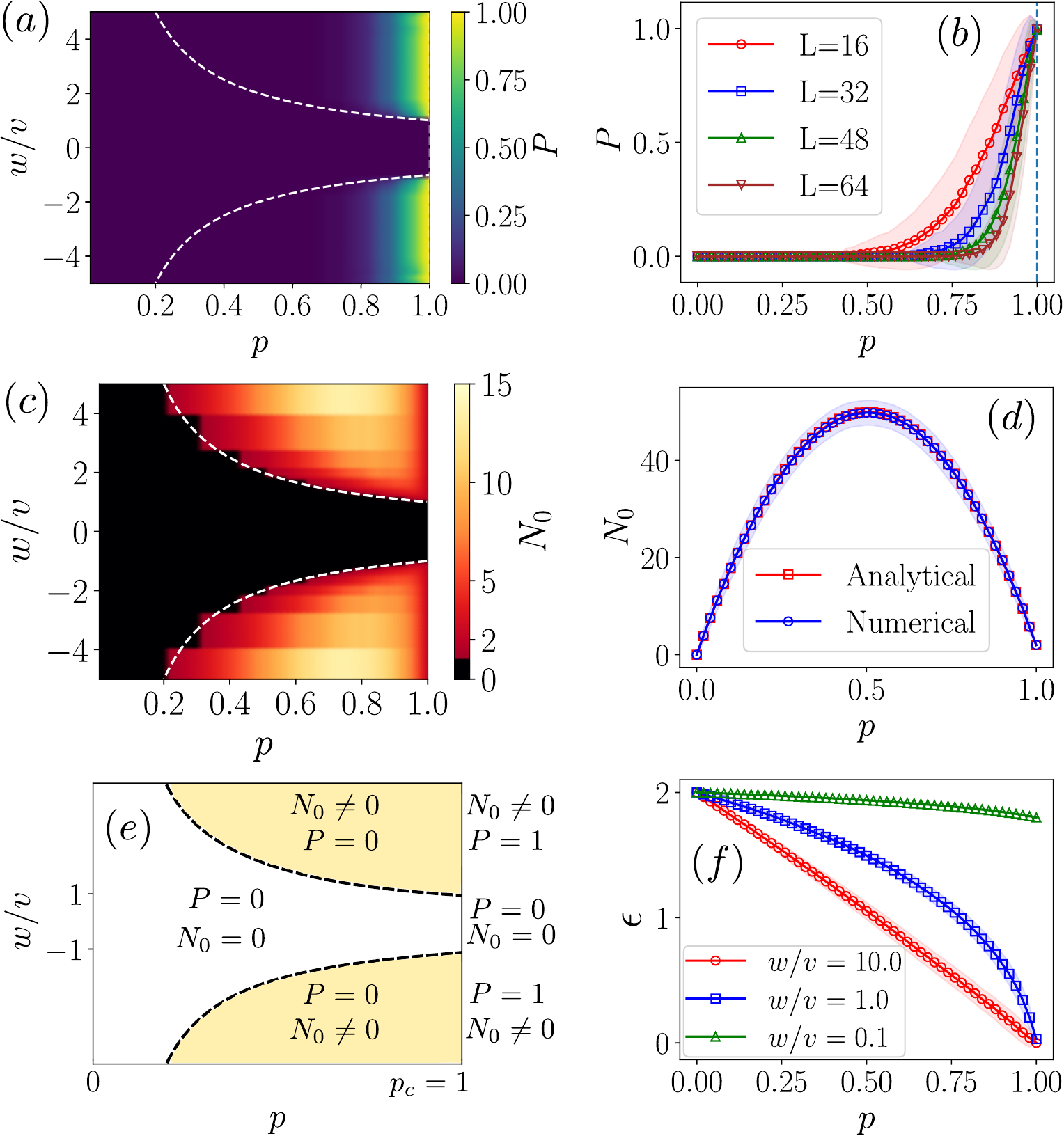}
    \caption{(a) Polarization ($P$) as a function of $w/v$ and $p$ for a bond-diluted SR-SSH chain (see Eqs.~\eqref{eq_Hsr} and~\eqref{eq_psr}) with $L=64$. (b) $P$ as a function of $p$ for different sizes $L$ ($w/v=10.0$). (c) The number of zero-energy modes ($N_{0}$) as a function of $w/v$ and $p$ for $L=64$. (d) Behavior of $N_{0}$ with $p$ in the deep topological regime ($w/v=100.0$) for $L=100$. (e) Schematic phase diagram with the parameters $w/v$ and $p$. (f) Mean energy-gap ($\epsilon$) per cluster (see Eq.~\eqref{eq_gc}) in the units of $v$ (where $v=1$) as a function of $p$ for $L=100$ for different $w/v$. The dotted lines in (a), (c), (e) are the mean-field critical lines: $wp=\pm v$. $p_{c}=1$ is the classical percolation threshold. Configuration averaging is performed over $200$ realizations in (a, c) and over $1000$ realizations in (b, d, f). The error-bars are always denoted by shaded thickness of the curves.}
    \label{fig:2}
\end{figure}

{\it{Phase diagram:}}
At $p=1$ we restore a clean SSH chain (see Ref.~\cite{shen_2012,asboth_2016,souvik_2021}), which is topologically non-trivial with two zero-modes ($N_{0}=2$) when $|w|>|v|$ and is trivial when $|w|<|v|$. The polarization (see Supplemental Material (SM)~\cite{supp}) takes
$P=1$ and $P=0$ in the two phases respectively, indicating the existence of bulk-edge correspondence at $p=1$. As the chain is diluted (i.e.~$p<1$), we find that the polarization is always zero ($P \sim 0$) for both $|w|>|v|$ and $|w|<|v|$, as evident from Fig.~\ref{fig:2}(a). Importantly, with increasing system size, $P \rightarrow 1$ (in the topological regime) only at $p = 1$ (see Fig.~\ref{fig:2}(b)). This would suggest that any amount of percolation in the SR-SSH chain immediately renders it trivial. However, the number of zero-modes in the system shows a nontrivial behavior which we now discuss. 

Given a typical Kubo gap in the system is $\Delta E \sim 1/L$, we associate the number of zero modes $=N_{0}$ to be the number of single-particle eigen-energies $E_i$ which lie within the window $|E_i| < \Delta E$. We find that $N_{0}$ is identically zero when $|w|<|v|$ for all values of $p$ but can take {\it non-zero} values when $|w|>|v|$ even when $p<1$ (see Fig.~\ref{fig:2}(c)). This indicates a breakdown of bulk-edge correspondence in the percolating (i.e. when $p<1$) topological phase - where even though the individual clusters have robust zero modes, the overall system does not have a finite polarization. This implies that polarization is a truly global object, which requires end-to-end connectivity in the system. Interestingly, $N_{0}$ goes to zero at a much lower value of $p$ given by, $\sim |v/w|$ shown by a dashed line in Fig.~\ref{fig:2}(c). To estimate this competing energy scale, we construct an effective ``mean-field" Hamiltonian ($H_{{\rm{MF}}}$) where a translationally invariant model is considered where hopping strength between unit cells ($w$) is scaled by the probability of having the corresponding connectivity $p$ (see \cite{supp}). This is similar to coherent potential approximation common in study of alloys~\cite{soven_prb1967, Velicky_PR_1969}. Thus, the trivial to topological phase transition here happens at $|wp|=|v|$. Therefore, while the appearance of the robust zero modes is governed by $H_{\rm{MF}}$, the behavior of polarization is determined by the geometric transition. The robustness of these zero modes are against further intra-cluster weak energetic disorder that continues to preserve the BDI symmetry class and are perturbative compared to the bulk gap scale.

Geometrical properties of the percolating clusters determine the number of zero-modes deep in the topological phase (i.e., $|w| \gg |v|$) since every cluster hosts two ``edge modes". 
An upper bound on $N_0$ is $ \approx 2 \sum_{s=2}^{\infty} {\mathcal{N}}_{s} = 2 L p (1-p)$
where ${\mathcal{N}}_{s}$ is the number of clusters with $s$ unit-cells (see ~\cite{supp}). This is confirmed numerically in Fig.~\ref{fig:2}(d). Thus the following qualitative phase diagram emerges: the percolation transition ($p_{c}=1$) is captured by $P$ while $N_{0}$ follows the mean-field critical lines $wp=\pm v$  (see Fig.~\ref{fig:2}(e)). The approach to percolation transition criticality predicts interesting scaling for physical quantities. For instance, $P$ approaches $1$ near $p \rightarrow p_c$ as 
$(1-P) \sim L^{-2/\nu_{p}}$
where $\nu_{p}=1$ and is thus determined by percolation exponent (see \cite{supp}). Similarly, the mean energy gap ($\epsilon$) per cluster, defined as
\begin{equation}\label{eq_gc}
\epsilon = \frac{\sum_{s} \epsilon_{s} {\mathcal{N}}_{s}}{\sum_{s} {\mathcal{N}}_{s}},
\end{equation}
exhibits interesting behavior near $p_c$ for different values of $w,v$. Here, $\epsilon_{s}$ is the energy-gap in a cluster with $s$ unit cells. A typical cluster of size $s$, for $|w| \gg |v|$ has $\epsilon_{s} \sim \exp(- s/\lambda)$ because of hybridizing edge modes. For $|w| \ll |v|$, the clusters are trivial and gapped with $\epsilon_{s} \sim 2|v-w|$ and at $|w|=|v|$, $\epsilon_{s} \sim 1/s$. 
This results in a linear behavior of $\epsilon \rightarrow 0$ as $p \rightarrow 1$ at $|w| \gg |v|$ and a logarithmic dependence at $|w|=|v|$ (see \cite{supp}). All these results are numerically confirmed in Fig.~\ref{fig:2}(f). Thus, the SR-SSH chain shows the non-trivial role of percolation physics in a topological phase. We now discuss long-range chains where $p_c$ can be tuned below $1$.

{\it{Bond-diluted LR-SSH chain:}} 
Geometrically, for long-ranged bond-diluted  chains~\cite{schulman_1983,schulman_1986,gori17}, where the probability of having a bond between $n$-th and $(n+r)$-th site goes as $p_{n,n+r}=p/r^{\alpha}$ (where $\alpha>1$), the percolation transition is found to occur at $p_{c}=1$ for $\alpha>2$ and $p_{c}<1$ for  $1<\alpha<2$, respectively, thus differentiating the two situations within a renormalization group study \cite{schulman_1986, gori17}. A lower bound on $p_{c}$ is given by Bethe threshold ($p_{c}^{{\rm{Bethe}}}$)~\cite{schulman_1983,gori17} which is the value of $p$ where a site is coupled to at least one other site, i.e., $p_{c}^{{\rm{Bethe}}} \sum_{r} (2/r^{\alpha}) \sim 1$, which implies $p_{c}^{{\rm{Bethe}}} \sim 1/(2 {\Li}_{\alpha}(1))$, where ${\Li}_{\alpha} (x) =\sum_{r=1}^{\infty} (x^{r}/r^{\alpha})$. In order to study the role of topological quantum Hamiltonian on such a LR chain we study the following model:
\begin{multline}\label{eq_Hlr}
H=v \sum_{n=1}^{L} (c^{\dagger}_{n,A} c_{n,B} + c^{\dagger}_{n,B} c_{n,A}) \\ + \sum_{n=1}^{L-1} \sum_{r=1}^{L-n} w_{n,n+r} (c^{\dagger}_{n,B} c_{n+r,A} + c^{\dagger}_{n+r,A} c_{n,B}), 
\end{multline}
where the probability distribution
of the inter-cell hopping strength $w_{n,n+r}$ between $n$-th and $(n+r)$-th unit cell is
\begin{equation}\label{eq_plr1}
{\Tilde{\mathcal{P}}}_{{\rm{LR}}} (w_{n,n+r})= \frac{p}{r^{\alpha}} \delta (w_{n,n+r}-w) + \Big(1- \frac{p}{r^{\alpha}}\Big) \delta (w_{n,n+r}), 
\end{equation} 
where $w_{n,n+r}$ can assume the values $w~(\neq 0)$ and zero with the probabilities $p/r^{\alpha}$ and $(1-(p/r^{\alpha}))$, respectively where $\alpha>1$. The probability of having a bond between the unit cells decreases with the distance $r$, but the inter-cell hopping strength ($w$) remains constant. The corresponding $H_{\text{MF}}$ where the inter unit-cell hopping goes as $\sim \frac{wp}{r^\alpha}$ (see \cite{supp}) governs physics which is determined by the average energetics of the problem.

\begin{figure}
\includegraphics[width=1.0\columnwidth]{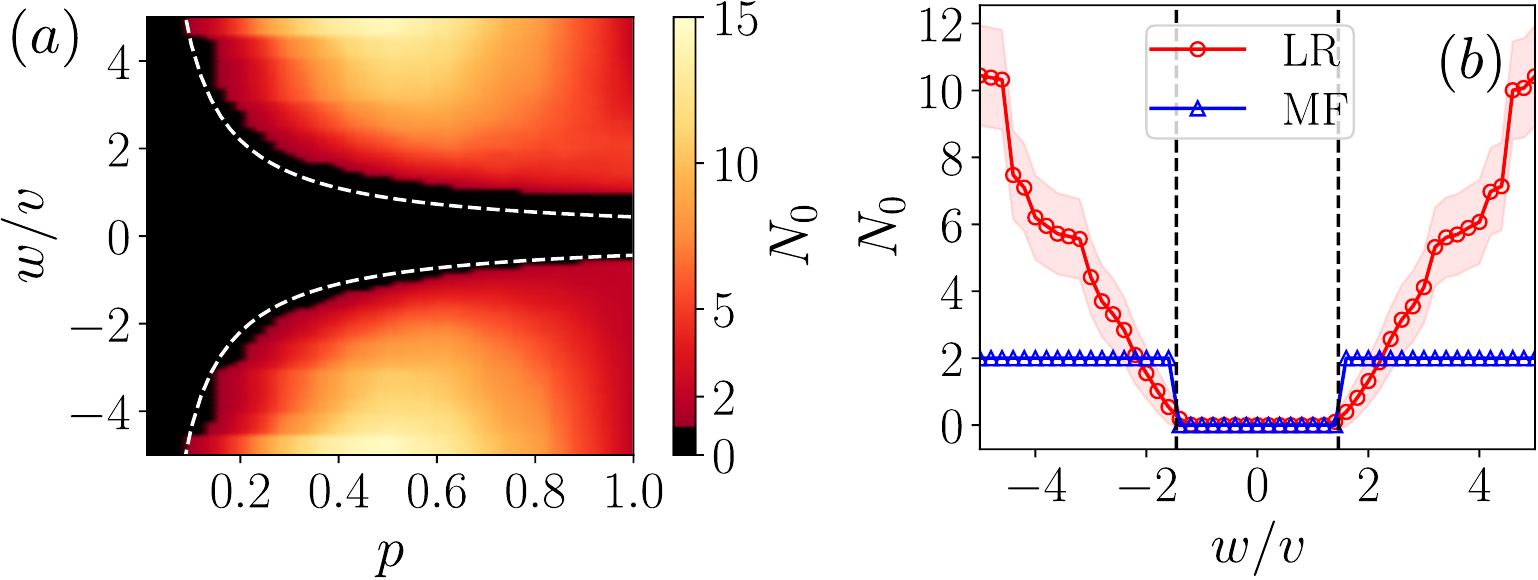}
\caption{(a) Number of zero-modes ($N_{0}$) as a function of $w/v$ and $p$ for Model-LR with $\alpha=1.6$ and $L=100$, (b) $N_{0}$ as a function of $w/v$ for LR and mean-field (MF) model with $\alpha=1.6$, $p=0.3$ and $L=100$. The dotted lines are the MF critical lines $wp=(-v)/(\Li_{\alpha}(-1))$ and $wp=(-v)/(\Li_{\alpha}(1))$. Configuration averaging is performed over $200$ realizations in (a) and over $1000$ realizations in (b).} \label{fig_lrz} \end{figure}
\begin{figure}
\includegraphics[width=1.0\columnwidth]{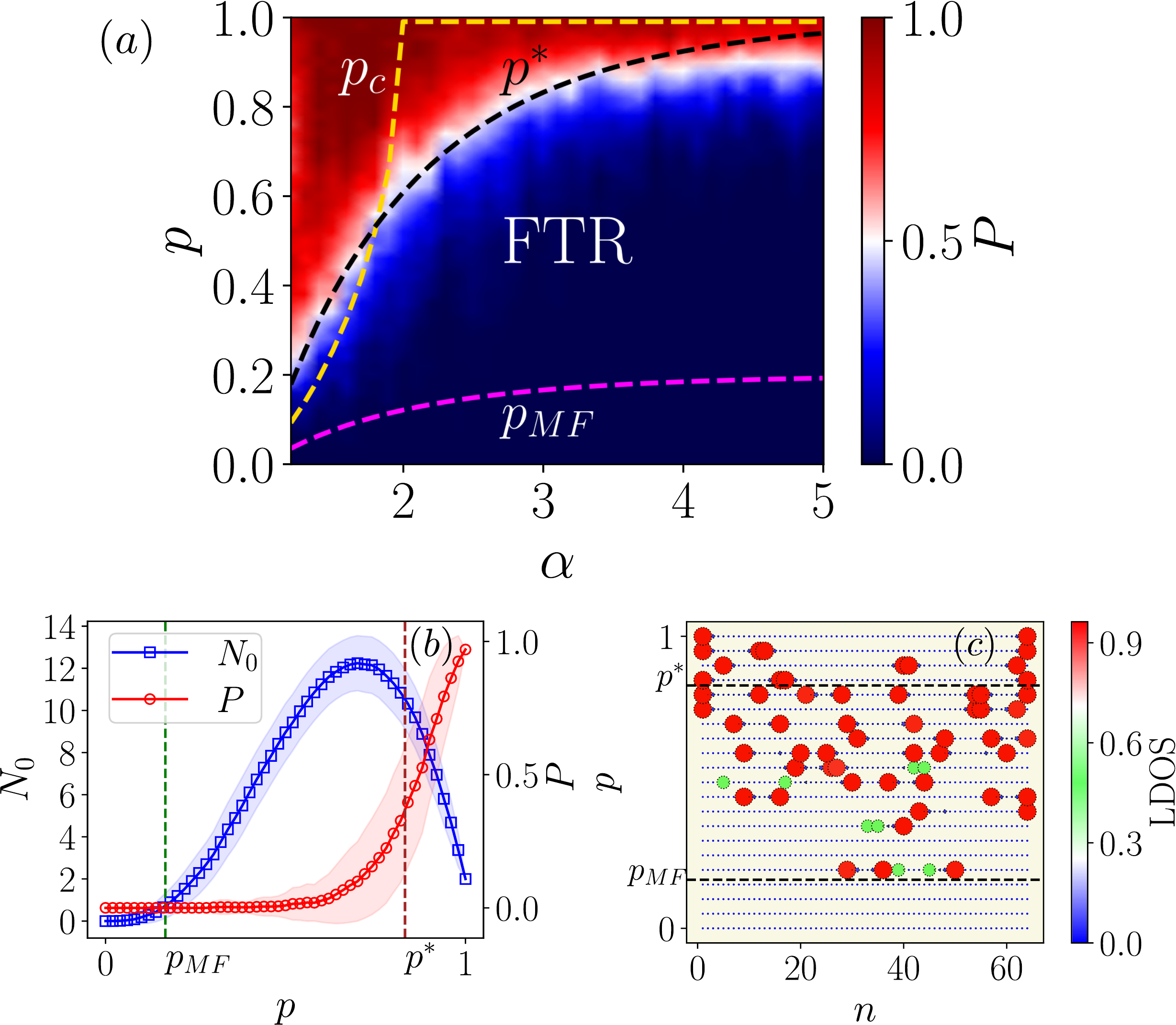}
\caption{(a) Polarization ($P$) as a function of $\alpha$ and $p$ for Model-LR with $|w|/|v|=5.0$ and $L=64$. (b) Number of zero-energy modes ($N_{0}$) and $P$ with $p$ for LR with $|w|/|v|=5.0$, $L=64$ and $\alpha=3.0$. (c) Local density of states (LDOS) of the zero-energy modes as a function of site index ($n$) and $p$ for LR with $|w|/|v|=5.0$ and $\alpha=3.0$. Here, $p_{c}$ is classical percolation threshold, $p^{*}=1/(\Li_{\alpha}(1))$ and $p_{MF}=|(-v)/(w\Li_{\alpha}(1))|$. Configuration averaging is performed over $200$ realizations in (a) and over $1000$ realizations in (b).} \label{fig_lrpalpha} \end{figure}

{\it{Results and Phase diagram:}} In LR system we calculate $N_{0}$ and find it is non-zero in the same parameter regime where the corresponding $H_{\text{MF}}$ is non-trivial (see \Fig{fig_lrz}(a) where the mean-field critical lines and $N_{0}$ are shown). The mean-field critical lines can be easily estimated to be $wp=(-v)/(\Li_{\alpha}(-1))$ and $wp=(-v)/(\Li_{\alpha}(1))$. This also corroborates with the behavior in the SR model in $\alpha \gg 2$ limit. The number of zero modes, however is dependent on the nature of the geometrical disorder -- as can be seen in \Fig{fig_lrz}(b) where Model-LR has significantly higher $N_0$ compared to the mean-field (MF) model. This is expected since for translationally invariant MF model $N_{0}=2$ in the topological regime; while in LR given a finite $p_c$ the system has larger number of disconnected clusters for the same value of $\alpha$ and $w/v$ (see \cite{supp}). We next investigate the behavior of polarization in this system. While in trivial regime ($|w|<|v|$) $P\sim 0$ irrespective of the value of $\alpha$ and $p$, deep in the 
topological regime the behavior of $P$ is quite unique. For $|w| \gg |v|$, when $\alpha \gg 2$, $P \sim 1$ only when $p\rightarrow 1$ consistent with SR system. However, as $\alpha$ is reduced, we find that $P$ undergoes a transition from $1$ to zero, at neither the mean-field value {\it nor} the geometrical percolation transition. The transition happens at a different crossover scale $p$ we label $p^*$. This divides the complete phase space into three distinct regions: for $p>p^*$, where both $P\sim 1, N_{0} \neq 0$ and thus the system is ``topological" even while percolation disorder is present. For $p<p_{\text{MF}}$ (where $p_{MF}=|(-v)/(w\Li_{\alpha}(1))|$) both $P$ and $N_{0}$ are $zero$ signalling a trivial phase. In between for $p>p_{\text{MF}}$ and $p<p^*$  while the system has a number of zero modes (coming from underlying topological clusters), $P\sim 0$ signalling an absence of any global topological response (see \Fig{fig_lrpalpha}(a),(b)). We dub this regime a Fractured Topological Region (FTR). Study of local density of states for zero energy modes [LDOS~$(n)=\sum_{i=1}^{N_{0}} |\langle n | \psi_{i} \rangle|^2$ , $n$ is site index] confirms such modes localized at the ``edges" of different clusters in FTR (see \Fig{fig_lrpalpha}(c)). The fractured region is thus characterized by an extensive number of zero modes (see \cite{supp}), determined by a combination of topological and geometrical properties - before the system eventually transits to a trivial phase. However, for $p>p^*$ the system undergoes a transition from topological to trivial phase as $w/v$ is tuned at a value determined by the $H_{\rm{MF}}$. This qualitatively points out the schematic phase diagrams shown in \Fig{fig:1} (b) and (c).

We note that while at $\alpha \gg 2$ value of $p^* = p_c \sim 1$, at small $\alpha$, $p^* \neq p_c$. Thus the polarization physics is not just  decided by the formation of a spanning cluster; in fact we find $p^*$ is broadly the value of $p$ when $\sum_{r=1}^{\infty} \frac{p}{r^\alpha} \sim 1$ (i.e. $p^* \sim ({\Li}_{\alpha}(1))^{-1}$) which is the probability when a site is connected to any site on one of its sides. Thus $p^*$ represents end-to-end connectivity between the two ends of the lattice which is $\rightarrow 1$ as $\alpha \gg 2$ thus coinciding with the geometrical $p_c$. In another  model where $w_{n,n+r}$ has the following probability distribution ($p_c=0$)
\begin{equation}\label{eq_plr2}
{\Tilde{\mathcal{P}}}_{{\rm{LR2}}} (w_{n,n+r})= p \delta \left( w_{n,n+r}-\frac{w}{r^{\alpha}} \right) + \left( 1- p \right) \delta (w_{n,n+r}),
\end{equation}
another $p^*$ emerges due to different percolation properties of the links (see \cite{supp}), even though the $p_{\text{MF}}$ remains the same as before. While our qualitative argument shows that the end-to-end connectivity gets established at $p^{*}$ we don't find any signatures of a thermodynamic phase transition for the electronic system. This is not unexpected since the electronic problem still resides on a highly disordered landscape. It will be interesting to find classical signatures of $p^*$ and how they are different from $p_c$.

{\it Discussion:} It is pertinent to note that while in a translationally symmetric Hamiltonian, a topological and trivial phase are separated by a critical point where both the topological invariant and its boundary manifestations undergo concurrent transitions - the same physics does not occur under a geometrical disorder. As a topological phase is punctured (here using a percolation protocol), it is expected that eventually the topological phase gives way to a trivial phase. However, our study shows that these two well-defined limits are separated by a large region where bulk-boundary correspondence breaks down. We also point out that even though we maintain the symmetry class of the underlying parent topological phase; these different regimes which seem to be separated out by distinct energy scales or geometrical properties are not separated via a critical point in the sense of a diverging length scale. The behavior of quantities such as gap, fidelity susceptibility, IPR and entanglement entropy suggest that all these phases and their crossover boundaries have localized states near Fermi energy (see \cite{supp}). In this sense, these transitions are also distinct from studies of perturbative disorder such as Anderson/hopping disorder where no geometrical phase transitions occur \cite{Perez_PRB_19,hsu_2020,shi_2021,liu2022topological}. We also point out that polarization and its real space equivalent as a topological response has been rigorously investigated for one dimensional topological insulators in both clean and disordered settings \cite{Perez_PRB_19,hsu_2020,shi_2021,liu2022topological}, however they haven't been discussed in context of geometrical transitions. While we have used $P$ to characterize the global topological response, it may be  useful to find other topological markers which may be better suited to capture the signatures of the percolation transition.

{\it{Outlook:}} While the nature of topological phases and their phase transitions are well understood in crystalline systems; an alternate method to trivialize a topological phase is via geometrical disorder. Unlike perturbative energetic disorder where a disorder energy scale is introduced - a geometrical disorder allows an interplay of both geometrical and Hamiltonian based phase transitions in the same system. We create toy models based on the simplest one-dimensional topological phases (SSH)  to delineate such interplay of geometrical and topological physics. In the SR model as the system is percolated, it creates macroscopic number of topological zero modes, even when the global $P\rightarrow 0$. In the LR model, where $p_c<1$, we find a new percolation scale $p^{*}$ above which polarization remains large, and below which we find a fractured topological region - where zero modes appear even in absence of global topological order. 

The experimental detection of signatures of percolation transition in physical observables like polarization for bond-diluted topological systems is an interesting open question. Experimental observation of the SSH topological phase in a Rydberg system~\cite{de2019observation} and a concrete proposal to study glassy physics in the same setup \cite{LiPRL2021} offer an exciting platform to realize this physics.  While our this work focuses on the one dimensional system, generalization of this study to higher dimensions and topologically ordered phases is a natural and exciting future direction. While a few such studies focusing on Chern insulators have appeared which show interesting hints of geometrical criticality \cite{Sahlberg_PRR_2020, Ivaki_PRR_2020}, it will be interesting to see if there is a general framework which opens up a new class of unique phases and phase transitions which are associated with just the geometrical properties of the system.

{\it{Acknowledgements:}}
We reminisce with love the memory of Prof. Amit Dutta with whom we started the discussions in this project. We acknowledge fruitful discussions with Diptarka Das, Deepak Dhar and Sumilan Banerjee. S.M. acknowledges support from PMRF Fellowship and S.P. acknowledges funding from IIT Kanpur Institute Fellowship. AA acknowledges support from IITK Initiation Grant (IITK/PHY/2022010). Numerical calculations were performed on the workstation {\it Wigner} at IITK.

\bibliography{ref_sshperc}









\newpage
\setcounter{equation}{0}
\makeatletter
\renewcommand{\theequation}{S\arabic{equation}}
\renewcommand{\thefigure}{S\arabic{figure}}

\onecolumngrid

\begin{center}
	\textbf{\large Supplemental Material to ``Percolation Transition in a Topological Phase"}
\end{center}

\vspace{\columnsep}
\vspace{\columnsep}

\twocolumngrid

\section{Polarization}
\label{app_pol}
In this section, we discuss the method used to compute polarization ($P$) for SSH chain. Let us consider the position operator
\begin{equation}\label{eq_pos_op}
\hat{X} = \sum_{n=1}^{L} (n-1) (c^{\dagger}_{n,A} c_{n,A} + c^{\dagger}_{n,B} c_{n,B}) ,
\end{equation}
and the projection operator
\begin{equation}\label{eq_projection}
\hat{\Pi} = \sum_{i=1}^{L} | \psi_{i} \rangle \langle \psi_{i} | ,
\end{equation}
where $| \psi_{i} \rangle$ is the eigenstate corresponding to $i$-th single-particle eigen-energy (where $2L$ eigen-energies are arranged in ascending order). The polarization~\cite{resta_1998} at the half-filling is then defined as
\begin{equation}\label{eq_pol}
P = {\im} \left[ {\Tr} \left[ \ln \left(\hat{\Pi} \exp \left(i \frac{2\pi \hat{X}}{L} \right) \hat{\Pi} \right) \right] \right] \text{~~modulo $(2\pi)$}.
\end{equation}

\section{Percolated SR-SSH Chain}

\subsection{Effective mean-field Hamiltonian}\label{app_mfsr}

In a bond-diluted SR-SSH chain, the mean value of inter-cell hopping strength ($w_{n,n+1}$) is ${\overline{w}}_{n,n+1}=wp$.  
We construct translationally invariant effective mean-field Hamiltonian ($H_{{\rm{MF}}}$) by replacing $w_{n,n+1}$ with its mean value $wp$. This is equivalent to averaging the Hamiltonian directly over multiple configurations. Thus, $H_{{\rm{MF}}}$ turns out to be
\begin{align}\label{eq_Hmfsr}
H_{{\rm{MF}}}=v \sum_{n=1}^{L} (c^{\dagger}_{n,A} c_{n,B} + c^{\dagger}_{n,B} c_{n,A}) \nonumber \\ 
+ \sum_{n=1}^{L-1} wp (c^{\dagger}_{n,B} c_{n+1,A} + c^{\dagger}_{n+1,A} c_{n,B}) .
\end{align}
In the periodic boundary condition, $H_{{\rm{MF}}}$ can be recast as
\begin{equation}\label{eq_kmfsr}
H_{{\rm{MF}}}= \sum_{k} \begin{pmatrix}
c^{\dagger}_{k,A} & c^{\dagger}_{k,B}
\end{pmatrix}
H_{k}
\begin{pmatrix}
c_{k,A} \\
c_{k,B}
\end{pmatrix}
.
\end{equation}
The Hamiltonian $H_{k}$ for the mode with momentum $k \in [-\pi,\pi]$ is given by
\begin{equation}\label{eq_hkmfsr}
H_{k}= (v + w p \cos(k) ) \sigma_{x} - w p \sin(k) \sigma_{y},
\end{equation}
where $\sigma_{x}$ and $\sigma_{y}$ are Pauli matrices.
Thus, the critical lines of $H_{\rm{MF}}$ are located at $wp=\pm v$. When $|wp|>|v|$, $H_{\rm{MF}}$ is non-trivial (polarization $P=1$ and number of zero-energy modes $N_{0}=2$)  and it is trivial ($P=0$, $N_{0}=0$) when $|wp|<|v|$.

\subsection{Zero-energy modes}\label{app_edge}

In the deep topologically non-trivial limit ($|w| \gg |v|$) of a bond-diluted, thermodynamically large ($L \to \infty$) SR-SSH chain, the total number of zero-energy modes ($N_{0}$) can be approximated as twice the number of all the clusters with the number of unit cells $s\geq 2$; as each of these clusters has two ``edge-modes". It is important to note that the clusters with one unit cell or $s=1$ are trivial (as there is only intra-cell hopping and no inter-cell hopping) and thus cannot have edge-modes. If ${\mathcal{N}}_{s}$ is the number of clusters with $s$ unit cells, then
\begin{equation}\label{eq_nzero}
N_{0} \approx 2 \sum_{s=2}^{\infty} {\mathcal{N}}_{s}  =2 L \sum_{s=2}^{\infty} n_{s} = 2 L p(1-p) ,
\end{equation}
where the number density of clusters~\cite{reynolds_1977,christensen_2002} with $s$ unit cells is given by $n_{s}= {\mathcal{N}}_{s}/L = (1-p)^2 p^{s-1}$.

\subsection{Finite-size scaling of polarization}\label{app_pscaling}

\begin{figure}
\centering
\includegraphics[width=1.0\linewidth]{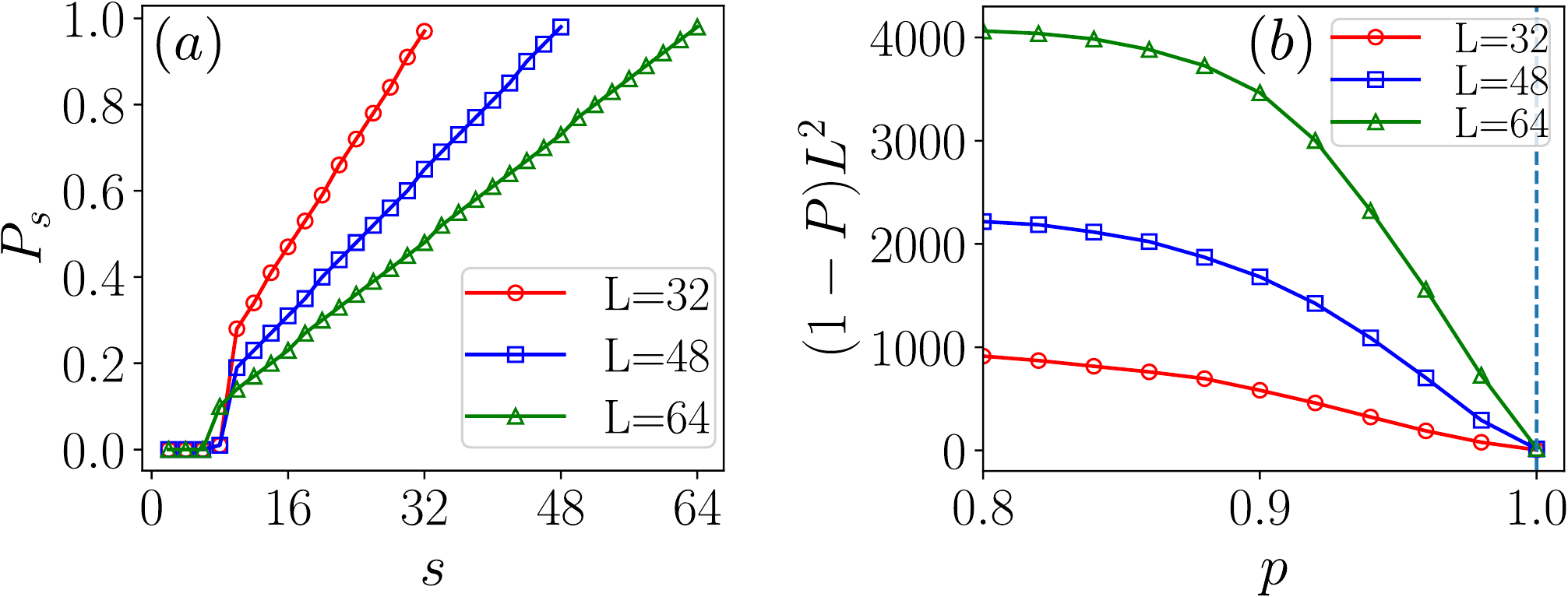}
\caption{(a) Polarization ($P_{s}$) due to a cluster with $s$ unit cells as a function of $s$ in a short-range (SR) SSH chain with $|w| \gg |v|$ and for different system-sizes $L$. (b) $(1-P)L^2$ as a function of $p$ in a bond-diluted SR-SSH chain with $w/v=10.0$.} \label{fig_psc} \end{figure}

In the deep topological phase ($|w|\gg |v|$), the clusters with more than one unit cell ($s \geq 2$) contribute $P_{s} \sim s/L$ (see Fig.~\ref{fig_psc}(a)) to the polarization (as the clusters with one unit cell or $s=1$ are trivial and cannot contribute to polarization). Thus, polarization $P$ of the chain can be expressed as
\begin{equation}\label{eq_pol_sum}
P \approx \sum_{s=2}^{\infty} P_{s}~{\mathcal{N}}_{s} \approx \sum_{s=2}^{\infty} \frac{s}{L} {\mathcal{N}}_{s} = 1- (1-p)^2,
\end{equation}
where ${\mathcal{N}}_{s}/L= (1-p)^2 p^{s-1}$. As the geometrical correlation length scales as $\xi_{p} \sim |p_{c}-p|^{-\nu_{p}}$, Eq.~\eqref{eq_pol_sum} can be recast as $(1-P) \sim \xi_{p}^{-2/\nu_{p}}$.
For a finite-sized system at $p \rightarrow p_{c}=1$, $\xi_{p} \gg L$ and thus $(1-P)$ scales as
\begin{equation}\label{eq_pol_finite}
(1-P) \sim L^{-2/\nu_{p}} \sim L^{-2} \text{~~~at $p \rightarrow 1$},
\end{equation}
as the percolation exponent $\nu_{p}=1$~\cite{reynolds_1977,christensen_2002}. The scaling of $P$ in Eq.~\eqref{eq_pol_finite} is also numerically confirmed (see Fig.~\ref{fig_psc}(b)).

\subsection{Mean energy-gap per cluster}\label{app_gap}
Now we calculate mean-energy-gap per cluster ($\epsilon$) for three different situations, namely $|w| \gg |v|$, $|w|=|v|$ and $|w| \ll |v|$.
\subsubsection{For $|w| \gg |v|$ limit}\label{app_nontrivial}
In this limit, the energy-gap between the ``edge-modes" in a cluster with $s\geq 2$ number of unit cells is $\epsilon_{s} \approx A \exp(- s/\lambda)$ (where $\lambda \sim 1/\ln(|w|/|v|)$ and $A$ is a constant). On the other hand, the clusters with one unit cell ($s=1$) are gapped (i.e. $\epsilon_{s=1} = 2|v|$). Therefore, it turns out that
\begin{equation}\label{eq_gcnts}
\epsilon = \frac{\sum_{s=1}^{\infty} \epsilon_{s}{\mathcal{N}}_{s}}{\sum_{s=1}^{\infty} {\mathcal{N}}_{s}} \approx \Big( 1-p \Big) \Big( 2 |v| + \frac{p \exp(-2/\lambda)}{\left( 1 - p \exp(-1/\lambda) \right)} \Big),
\end{equation}
where ${\mathcal{N}}_{s}/L= (1-p)^2 p^{s-1}$. As $\lambda \to 0$ when $|w| \gg |v|$, Eq.~\eqref{eq_gcnts} reduces to \begin{equation}\label{eq_gcntf}
\epsilon \approx 2 |v| (1-p),
\end{equation}
indicating a linear decrease in $\epsilon$ with $p$.

\subsubsection{For $|w|=|v|$ limit}\label{app_critical}

In this limit, for the clusters with $s \geq 2$, the energy-gap $\epsilon_{s} \approx b/s$ (where $b$ is a constant) and for the clusters with $s=1$, $\epsilon_{s} = 2|v|$. Thus, we obtain
\begin{equation}\label{eq_gccrs}
\epsilon \approx \Big( 1 - p \Big) \Big( 2 |v| + \frac{b{\mathcal{M}}}{p} \Big),
\end{equation}
where ${\mathcal{M}} =\sum_{s=2}^{\infty} ~(p^{s}/s) = - \ln(1-p) - p$. 
Therefore, it turns out that
\begin{equation}\label{eq_gccrf}
\epsilon \approx \Big( 1 - p \Big) \left( 2 |v| - \frac{b}{p} \ln(1-p) -b \right),
\end{equation} signalling a logarithmic dependence of $\epsilon$ with $p$.

\subsubsection{For $|w| \ll |v|$ limit}\label{app_trivial}
For $|w| \ll |v|$, all the clusters are trivial and gapped with the energy-gap $\epsilon_{s} \approx 2|v-w|$. Therefore, $\epsilon \approx 2 |v-w|$.

\section{Percolated LR-SSH Chain}

\subsection{Mean-field Hamiltonian}\label{app_mflr}
In the bond-diluted LR-SSH chain, the mean value of inter-cell hopping strength is ${\overline{w}}_{n,n+r} = wp/r^{\alpha}$ and thus the mean-field Hamiltonian is given by
\begin{align}\label{eq_Hmflr}
H_{{\rm{MF}}}=v \sum_{n=1}^{L} (c^{\dagger}_{n,A} c_{n,B} + c^{\dagger}_{n,B} c_{n,A}) \nonumber \\ 
+ \sum_{n=1}^{L-1} \sum_{r=1}^{L-n} \frac{wp}{r^{\alpha}} (c^{\dagger}_{n,B} c_{n+r,A} + c^{\dagger}_{n+r,A} c_{n,B}) .
\end{align}
The Hamiltonian $H_{k}$ for the mode with momentum $k \in [-\pi,\pi]$ then assumes the form
\begin{equation}\label{eq_hkmflr}
H_{k}= (v + w p {\real} ({\Li}_{\alpha}(e^{-ik})) ) \sigma_{x} + w p {\im} ({\Li}_{\alpha}(e^{-ik})) \sigma_{y},
\end{equation}
where ${\Li}_{\alpha}(x)= \sum_{r=1}^{\infty} (x^{r}/r^{\alpha})$. 
Thus, critical lines of $H_{\rm{MF}}$ are $wp=(-v)/({\Li}_{\alpha}(-1))$ and $wp=(-v)/({\Li}_{\alpha}(1))$. When $wp > (-v)/({\Li}_{\alpha}(-1))$ and $wp < (-v)/({\Li}_{\alpha}(1))$, $H_{\rm{MF}}$ is non-trivial ($P=1$ and $N_{0}=2$). On the other hand, $H_{\rm{MF}}$ is trivial ($P=0$, $N_{0}=0$) when $(-v)/({\Li}_{\alpha}(1) < wp < (-v)/({\Li}_{\alpha}(-1))$.

It is noteworthy that for Model-LR2, $H_{\rm{MF}}$  is same as that for LR with the same set of parameters since for both the models, ${\overline{w}}_{n,n+r} = wp/r^{\alpha}.$

\subsection{Zero-modes, polarization and Model-LR2}\label{app_lr2}

\begin{figure}
\centering
\includegraphics[width=1.0\columnwidth]{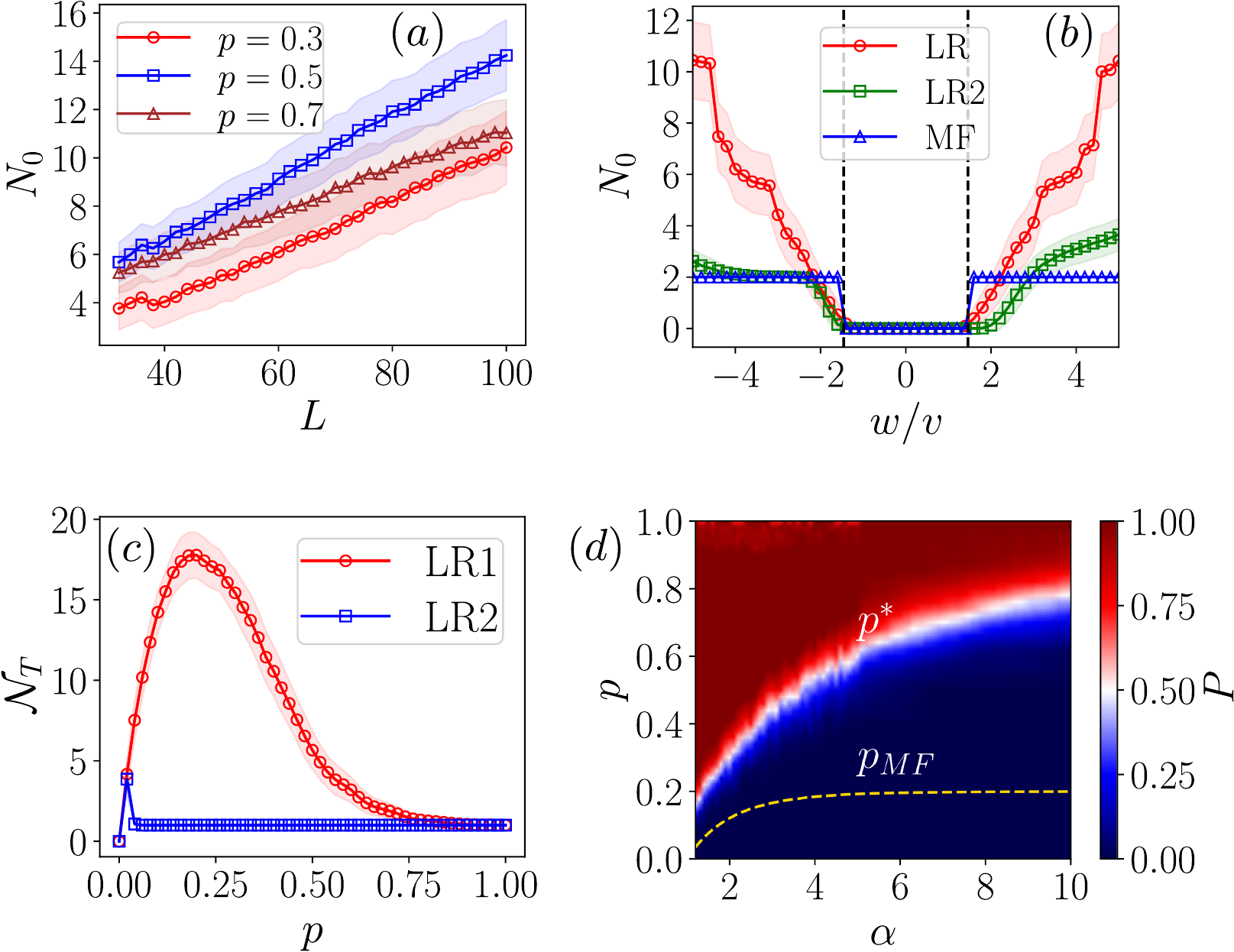}
\caption{(a) Number of zero-energy modes ($N_{0}$) as a function of $L$ for Model-LR with $\alpha=1.6$ and $|w|/|v|=5.0$. (b) $N_{0}$ as a function of $w/v$ for Model-LR, LR2 and mean-field model (MF) with $\alpha=1.6$, $L=100$ and $p=0.3$. (c) Number of clusters ($\mathcal{N}_{T}$) of size $s\geq 2$ as a function of $p$ for Model-LR and LR2 with $L=100$ and $\alpha=1.6$. (d) Polarization ($P$) as a function of $\alpha$ and $p$ for Model-LR2 with $L=64$, $|w|/|v|=5.0$. The dotted lines in (b) are mean-field critical lines $wp=(-v)/({\Li}_{\alpha}(-1))$ and $wp=(-v)/({\Li}_{\alpha}(1))$. In (d), $p_{MF}=|(-v)/(w {\Li}_{\alpha}(1))|$. Configuration averaging is performed over $1000$ realizations in (a, b, c) and over $200$ realizations in (d).}\label{fig_lrb}
\end{figure}

\subsubsection{Zero-energy modes} The number of zero-energy modes ($N_{0}$) in the deep topological regime ($|w|\gg |v|$) of Model-LR is extensive in system-size $L$, i.e. $N_{0} \sim L$ (see Fig.~\ref{fig_lrb}(a)). On the other hand, for Model-LR2 $N_{0}$ remains close to $2$ when the corresponding $H_{\rm{MF}}$ is non-trivial (see Fig.~\ref{fig_lrb}(b)). This can  be explained as follows: at a finite $p$ (i.e. $0<p<1$), the number of clusters ($\mathcal{N}_{T}$) having zero-energy modes for LR is much greater than that for LR2 with the same set of parameters ($p,\alpha,L$), as evident from Fig.~\ref{fig_lrb}(c). It is noteworthy that $\mathcal{N}_{T}$ for Model-LR and LR2 in the limit $|w| \gg |v|$ can be obtained from the total number of clusters with more than one unit cell ($s\geq 2$), i.e. ${\mathcal{N}_{T}} = \sum_{s=2}^{\infty} {\mathcal{N}}_{s}$.

\subsubsection{Polarization} In the deep topological regime ($|w|\gg |v|$) of Model-LR2, the polarization ($P$) shows a transition from {\it zero} to $P \sim 1$ at a value of $p=p^{*}$, which is different from both $p_{MF}$ (obtained from mean-field critical lines) and the geometrical $p_{c}=0$, as can be seen Fig.~\ref{fig_lrb}(d). In the large $\alpha$ limit, $p^{*}$ approaches $1$, which is consistent with the short-range limit. However, as $\alpha$ is decreased, $p^{*}$ decreases from $1$ due to the introduction of long-range inter-cell couplings. We find that $p^*$ for LR2 is smaller than the Model-LR discussed in the main text since
due to higher network connectivity the topological phase is realized for smaller values of $p$.

\subsection{Additional numerical results}\label{app_gap_ent}

\begin{figure}
\centering
\includegraphics[width=1.0\columnwidth]{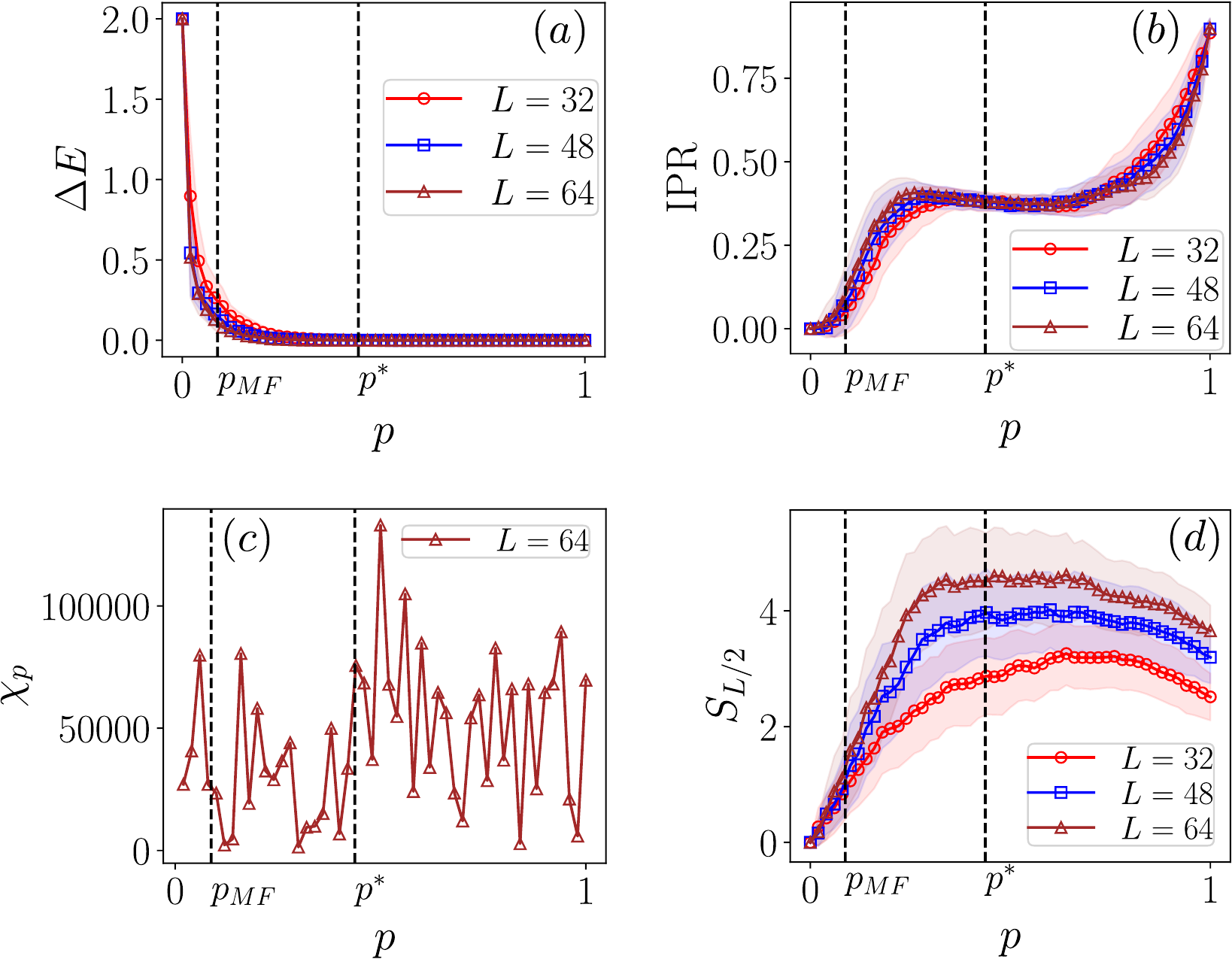}
\caption{(a) Energy-gap ($\Delta E$), (b) inverse participation ratio (IPR), (c) fidelity susceptibility
$\chi_{p}=\big|\frac{\partial^{2} E_{g}}{\partial p^{2}} \big|$ (where $E_{g}$ is the ground state energy) and (d) bipartite entanglement entropy $S_{L/2}$
 as the functions of $p$ for bond-diluted long-range (LR) SSH chain with $\alpha=1.6$ and $|w|/|v|=5.0$. Here, $p_{MF}=|(-v)/(w {\Li}_{\alpha} (1))|$ and $p^{*}=1/({\Li}_{\alpha} (1))$. In all the plots, configuration averaging is performed over $200$ realizations.}\label{fig_gap_ipr}
\end{figure}

\subsubsection{Energy-gap} 

From the study of the energy-gap ($\Delta E$) we find that the bond-diluted LR-SSH chain in the limit $|w|\gg |v|$ is gapped when $p<p_{MF}$ (i.e., when $H_{MF}$ is trivial) where $p_{MF}=|(-v)/(w {\Li}_{\alpha}(1))|$. On the other hand, for $p>p_{MF}$, $\Delta E \sim 0$ (see Fig.~\ref{fig_gap_ipr}(a)) due to the existence of significant number of zero-energy modes.

\subsubsection{Inverse participation ratio} 

Inverse participation ratio (IPR) of the zero-energy modes is defined as,
\begin{equation}\label{eq_ipr}
{\rm{IPR}}=  \frac{1}{N_{0}} \sum_{i=1}^{N_{0}} \sum_{n=1}^{2L} |\langle n | \psi_{i} \rangle|^4 , 
\end{equation}
where $i$ runs over all the zero-energy eigenstates and $n$ is the site index. For a bond-diluted LR-SSH chain in the limit $|w|\gg |v|$, we find that IPR of the zero-energy modes is independent of $L$ (see Fig.~\ref{fig_gap_ipr}(b)). This result indicates that the zero-energy modes are localized at the ``edges" of different clusters.

\subsubsection{Fidelity susceptibility} A version of fidelity susceptibility can be defined as $\chi_{p}=\big| \frac{\partial^{2} E_{g}}{\partial p^{2}} \big|$ where $E_{g}$ is the ground state energy. However, $\chi_{p}$ does not show any significant feature at $p^{*}$ and $p_{MF}$ (see Fig.~\ref{fig_gap_ipr}(c)).

\subsubsection{Bipartite entanglement entropy} Bipartite entanglement entropy $S_{L/2}$ for bond-diluted LR-SSH chain in the limit $|w|\gg |v|$ is independent of $L$ when $p<p_{MF}$. 
This result is consistent with the area-law of entanglement in a gapped phase. However, for $p>p_{MF}$, $S_{L/2}$ increases with $L$ (see Fig.~\ref{fig_gap_ipr}(d)), which occurs due to the rise in the number of zero-energy modes with $L$ (i.e. $N_{0} \sim L$).



\end{document}